\documentclass[twoside]{dis09}
\usepackage[latin1]{inputenc}
\usepackage[dvips]{graphicx,epsfig,color}
\usepackage{wrapfig,rotating}
\usepackage{amssymb,amsmath,array}
\graphicspath{{./pics/}}

\begin{document}

\newcommand{\pb}{\mathrm{pb}}
\newcommand{\fb}{\mathrm{fb}}
\newcommand{\GeV}{\mathrm{GeV}}
\newcommand{\TeV}{\mathrm{TeV}}
\newcommand{\mausbox}[1]{\colorbox[cmyk]{0,0,0,0.2}{#1}}
\def\muf{{\mu^{}_f}}
\def\mur{{\mu^{}_r}}
\def\mt{{m_t}}
\newcommand{\msbar}{\ensuremath{\overline{\mbox{MS}}}}

\title{
\textnormal{\normalsize
            \phantom{m}\\[-15mm]
            DESY-09-104\phantom{MMMMMMMMMMMMMMMMMMMMMMMMMMMMMMMMMI}\\[0mm]
            SFB/CPP-09-61\phantom{MMMMMMMMMMMMMMMMMMMMMMMMMMMMMMMM}\\[0mm]
            HU-EP-09/31\phantom{MMMMMMMMMMMMMMMMMMMMMMMMMMMMMMMMMI}\\[2mm]
            }
New results for $t\bar{t}$ production at hadron colliders}

\author{
U. Langenfeld$^1$, S. Moch$^1$, and P. Uwer$^2$
%
\thanks{Talk given by U. Langenfeld at the 
        XVII International Workshop on Deep-Inelastic Scattering and 
        Related Subjects, DIS 2009, 26-30 April 2009, Madrid, see Ref.~\cite{url}}
%
\vspace{.3cm}\\
%
1- DESY Zeuthen \\
Platanenallee 6, 15738  Zeuthen - Germany
%
\vspace{.1cm}\\
2-  Humboldt-Universität zu Berlin, Institut für Physik\\
 Newtonstr. 15, 12489 Berlin - Germany\\
}
\date{\today}
\maketitle

\begin{abstract}
We present new theoretical predictions for the $t\bar{t}$ production cross section
at NNLO at the Tevatron and the LHC. We discuss the scale uncertainty
and the errors due to the parton distribution functions (PDFs).
For the LHC, we present a fit formula for the pair production cross section
as a function of the center of mass energy
and we provide predictions for the pair production cross section
of a hypothetical heavy fourth generation quark $t'$.
\end{abstract}

\section{$t\bar{t}$ - Production at Tevatron and LHC}
The experimental measurements of the top mass $m_t$ and the $t\bar{t}$ production 
cross section have reached a relative accuracy of $0.75\%$~\cite{:2009ec} and
$9\%$~\cite{Lister:2008it}, respectively.
Therefore it is necessary to provide improved theoretical predictions for the total
cross section at the Tevatron and the LHC in perturbative QCD.

The total hadronic cross section for $t\bar{t}$ production depends
on the top mass $\mt$, the center of mass energy $s= E^2$, the factorisation scale $\mu_f$,
the renormalization scale $\mu_r$, and the PDF set and it is given by
\begin{equation}
 \sigma(s,m_t,\mu_r,\mu_f) = \sum_{i,j=g,q,\bar{q}} 
            f_{i/p}(\mu_f^2)\otimes f_{j/p}(\mu_f^2) \otimes \hat{\sigma}(m_t,\mu_r,\mu_f),
\end{equation}
where $f_{i/p}$ are the proton PDFs.
In the following, we discuss these dependencies.

We have updated the results from Refs.~\cite{Moch:2008qy,Moch:2008ai} as follows.
To obtain more reliable estimates of the scale uncertainty we have used the 
full dependence on $\mur$ \emph{and} $\muf$.
We have performed a consistent singlet - octet - decomposition when matching our NNLO 
threshold expansion at NLO.
Further corrections 
(electroweak contributions~\cite{Beenakker:1993yr,Bernreuther:2006vg,Kuhn:2006vh}, 
QCD bound state effects near threshold~\cite{Kiyo:2008bv}, and
new parton channels $qq$, $\bar{q}\bar{q}$, and $q_i\bar{q}_j$ for unlike quarks
opening at NNLO~\cite{Dittmaier:2007wz,Dittmaier:2008uj}
) are generally small and have been estimated.

As a new result we have studied the dependence of the $t\bar{t}$ production cross section 
on the definition of the mass parameter.
We have used the \msbar~mass scheme as an alternative mass description
exploiting the conversion relation between the pole mass $\mt$ and the \msbar~mass
$m(\mur)$~\cite{Gray:1990yh}.
We find that the convergence of the perturbation expansion through NNLO is improved
using the \msbar~mass.
This expansion has a considerably reduced scale dependence even at NLO.
The NLO and NNLO corrections in the \msbar~scheme are much smaller than 
the corresponding corrections in the pole mass scheme.
Therefore, we find good properties of convergence of the perturbation series.
From the measured $t\bar{t}$ cross section at the Tevatron~\cite{Abazov:2009ae}
we derive a \msbar~mass $\overline{m} = 160.0^{+3.3}_{-3.2}\,\GeV$,
which corresponds to a pole mass of $168.9^{+3.5}_{-3.4}\,\GeV$.
More details of this analysis are presented in~\cite{Langenfeld:2009wd}.
Throughout this article, we have chosen the PDF set CTEQ6.6~\cite{Nadolsky:2008zw}.
In Ref.~\cite{Langenfeld:2009wd}, results for the PDF set
MSTW NNLO 2008~\cite{Martin:2009iq} can be found.
The top mass is the pole mass and is set to the most recent 
value $m_t = 173\,\GeV$~\cite{:2009ec} if not otherwise stated.

\begin{figure}
\begin{minipage}{69mm}
\centering
\scalebox{0.85}{\includegraphics[%
          bbllx=77pt,bblly=541pt,bburx=298pt,bbury=767pt]{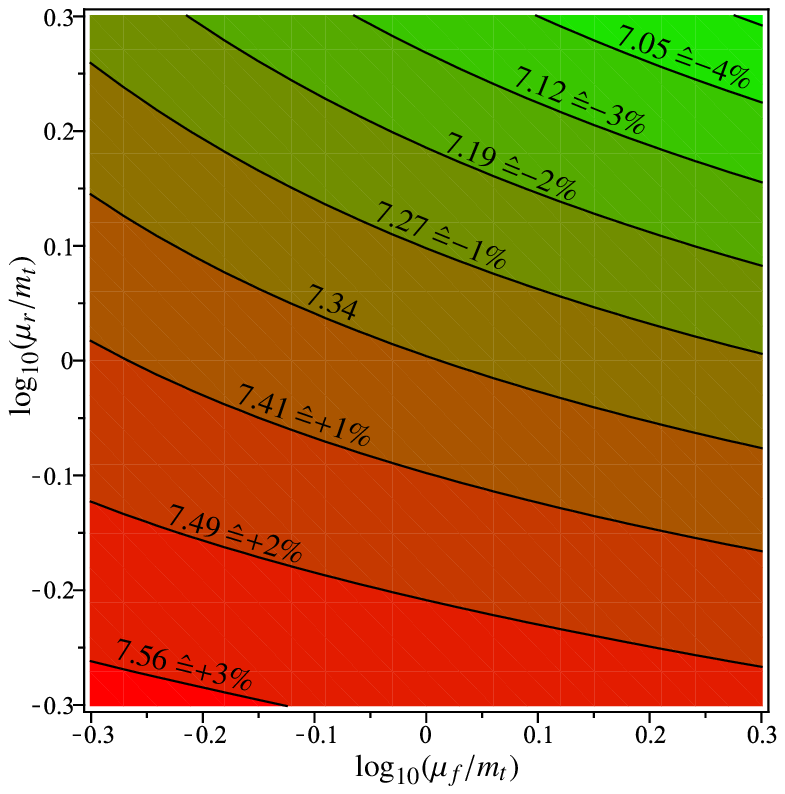}}
\end{minipage}
\begin{minipage}{69mm}
\centering
\scalebox{0.85}{\includegraphics[%
          bbllx=77pt,bblly=541pt,bburx=298pt,bbury=767pt]{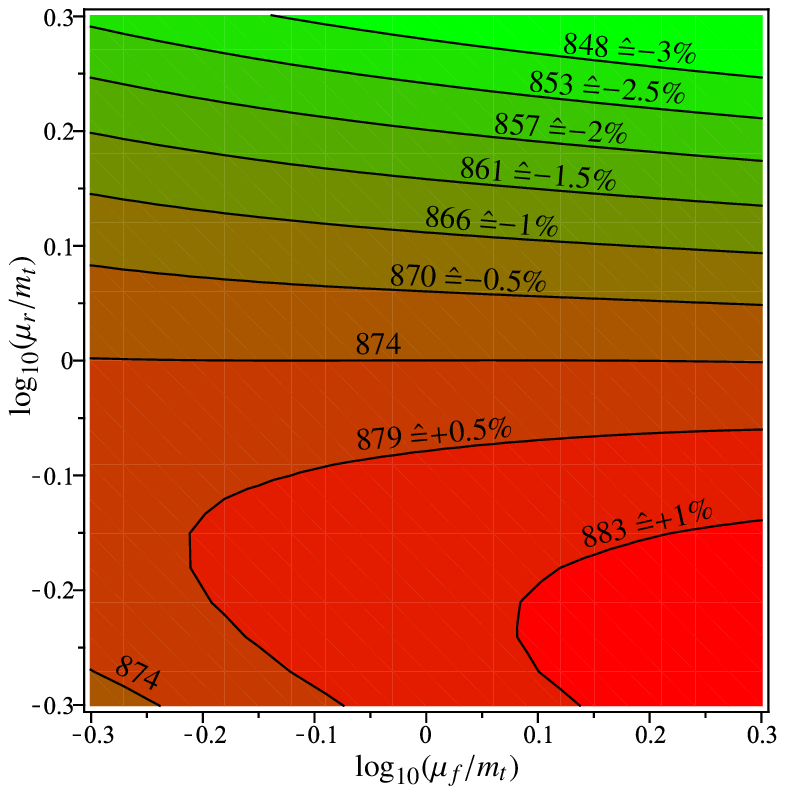}}
\end{minipage}

\begin{minipage}{140mm}
 \caption{
  \label{fig:totalXsection}
  Contour lines of the total hadronic cross section 
  from the independent variation of renormalization and factorization scale 
  $\mur$ and $\muf$ for the Tevatron with $\sqrt{s} = 1.96\,\TeV$ (left) 
  and the LHC with $\sqrt{s} = 14\,\TeV$ (right) with
  CTEQ6.6\cite{Nadolsky:2008zw}.
  The range of $\mur$ and $\muf$ corresponds to $\mur,\muf \in [\mt/2,2\mt]$.
}
\end{minipage}
\end{figure}

We have analysed the dependence of the cross section on the renormalisation \emph{and} 
factorisation scale.
In Fig.~\ref{fig:totalXsection}, we display the result for the Tevatron and the LHC.
At the Tevatron, the gradient is nearly parallel to the diagonal, and we find errors
of $-5\%$ at $(\muf,\mur)=(\mt/2,\mt/2)$ and $+3\%$ at $(2\mt,2\mt)$.
Likewise for the LHC, the scale uncertainty is about $1\%$ at about
$(2\mt,\mt/2)$ and $-4\%$ at $(2\mt,2\mt)$.
Note that in the case of the LHC, the cross section is not a monotonically decreasing 
function if $\mur=\muf$ as it is in the case of the Tevatron,
see Ref.~\cite{Langenfeld:2009wd} for details.

In Figs.~\ref{fig:teva:massdep} and~\ref{fig:lhc:massdep}, we show the mass dependence
of the total hadronic cross section for both colliders including the scale uncertainty
for $\mur = \muf \equiv \mu = \mt/2$ and $\mu =  2\mt$.

The pure PDF error $\Delta \mathcal{O}$ is given by
\begin{equation}
 \Delta \mathcal{O} = 
 \sqrt{\frac{1}{2}\sum_{k=1,n_{\mathrm{PDF}}} \big(\sigma_{k+} -\sigma_{k-}\big)^2},
\end{equation}
where $\Delta \mathcal{O}$ is determined from the variation of the cross section
with respect to the parameters of the PDF fit.
The PDF errors are added linearly. 
The result is presented in Fig.~\ref{fig:teva:pdferrormstw} for the Tevatron 
and in Fig.~\ref{fig:lhc:pdferrormstw} for the LHC.
We show for both colliders the NLO and NNLO cross sections together with their 
error bands.
This demonstrates the shrinking of the total error for the NNLO cross section.

Having discussed scale uncertainty and PDF error, we present our prediction
for the cross section at the Tevatron and the LHC.
To obtain a more conservative error bound,
we calculate the contribution of the scale uncertainty as
\begin{equation}
 \min_{\mu_r,\mu_f \in [m_t/2,2 m_t]}\sigma(\mu_r,\mu_f) \leq \sigma(\mu_r,\mu_f)
 \leq\max_{\mu_r,\mu_f \in [m_t/2,2 m_t]}\sigma(\mu_r,\mu_f).
\end{equation}
For $t\bar{t}$ production at the Tevatron, this definition changes nothing,
but for $t\bar{t}$ production at the LHC, the upper bound is shifted to
larger values by a few per cent. See also Fig.~\ref{fig:totalXsection} and 
the corresponding discussion. For the CTEQ6.6 PDF set and $\mt=173\,\GeV$
(pole mass), we arrive at
\begin{center}
\mausbox
{
\parbox{62mm}{
$\sigma(p\bar{p}\to t\bar{t}) = 7.34^{+0.23}_{-0.38}\,\pb$ @ Tevatron,\\[2mm]
$\sigma(p p\to t\bar{t}) = 874^{+14}_{-33}\,\pb$ @ LHC $14\,\TeV$.
}
}
\end{center}

\begin{figure}
\vspace{7mm}
\begin{minipage}{70mm}
\centering
\hspace*{-5mm}
\scalebox{0.28}{\includegraphics{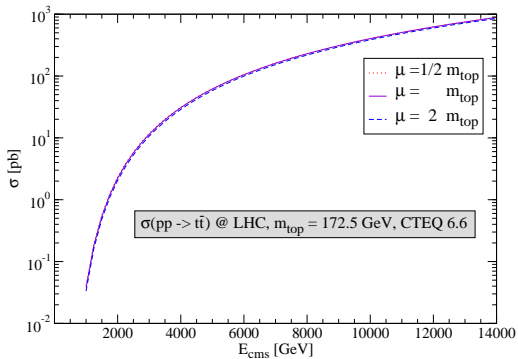}}
\end{minipage}
\hspace*{5mm}
\begin{minipage}{65mm}
{\caption{$t\bar{t}$ production at the LHC as a function
           of the center of mass energy $E$ for $m_t=172.5\,\GeV$
           and for three different scales $\mu = 1/2 m_t, m_t, 2 m_t$,
           see Eq.~(\ref{eq:fitcms}) and Tab.~\ref{tab:hadrofitcteq}
           for the parametrisation.
           \vspace*{25mm}}
\label{fig:lhc:cmsdep}}
\end{minipage}
\end{figure}

For the LHC, we have calculated the total hadronic cross section
as a function of the center of mass energy $E$ for a value of 
$m_t=172.5\,\GeV$ as used in ATLAS studies, see Fig.~\ref{fig:lhc:cmsdep}.
We parametrize the result using the ansatz
\begin{equation}
\label{eq:fitcms}
\sigma(E,\mu) = a + b x + c x^2
                    + d x\log\left(\frac{E}{\sqrt{s}}\right)
                    + e x\log^2\left(\frac{E}{\sqrt{s}}\right)
                    + f x^2\log\left(\frac{E}{\sqrt{s}}\right)
                    + g x^2\log^2\left(\frac{E}{\sqrt{s}}\right)
\end{equation}
with $x=E/\GeV$ and $\sqrt{s} = 14\,\TeV$. 
The numerical values for the coefficients $a,\ldots,g$ can be found in 
Tab.~\ref{tab:hadrofitcteq} for $\mu = m_t$, $m_t/2$, and $2 m_t$. 
The fit is valid for $3\, \TeV \leq E \leq 14 \,\TeV$ 
and has an accuracy of better than $0.05\%$ within this range.
This ansatz is justified by general limits for cross sections at high energies
and is consistent with unitarity.
Parametrisations of the total cross section as a function of $\mt$
can be found for various scenarios in Ref.~\cite{Langenfeld:2009wd}.

\begin{table}[ht!]
\renewcommand{\arraystretch}{1.25}
{\small
\centering
\begin{tabular}{|l|r|r|r|r|r|r|r|}
\hline
&\multicolumn{1}{c|}{$a[\times 10]$}&\multicolumn{1}{c|}{$b[\times10^{-1}]$}
&\multicolumn{1}{c|}{$c[\times10^{-5}]$}& \multicolumn{1}{c|}{$d[\times10^{-1}]$}&
\multicolumn{1}{c|}{$e[\times10^{-2}]$}&\multicolumn{1}{c|}{$f[\times10^{-5}]$}&
 \multicolumn{1}{c|}{$g[\times10^{-6}]$}\\[1mm]
\hline
$\sigma(\mu = m_t)$ &$    3.42553  $ &$   -5.12699  $ &$    4.09683  $ &$   -2.46892  $
&$   -3.93892  $ &$   -1.75175  $ &$    2.02029  $   \\[1mm]
  $\sigma(\mu = m_t/2)$ &$    3.20912  $ &$   -4.85885  $ &$    3.90541  $ &$   -2.33781  $ 
  &$   -3.71957  $ &$   -1.65930  $ &$    1.88132  $   \\[1mm]
  $\sigma(\mu = 2 m_t)$ &$    3.31748  $ &$   -4.76706  $ &$    3.82310  $ &$   -2.31392  $ 
  &$   -3.73054  $ &$   -1.60468  $ &$    1.74661  $  \\[1mm]

\hline
\end{tabular}
\caption{
Numerical values of the coefficients (in $\pb$) of  Eq.~\ref{eq:fitcms}
for $\mt = 172.5\,\GeV$ and the PDF set CTEQ6.6.
\label{tab:hadrofitcteq}
}}
\end{table}

\section{Predictions for $t'\bar{t}'$ Production at Tevatron and LHC}
\begin{figure}[ht!]
\begin{minipage}{69mm}
\centering
\hspace*{-5mm}
\scalebox{0.28}{\includegraphics{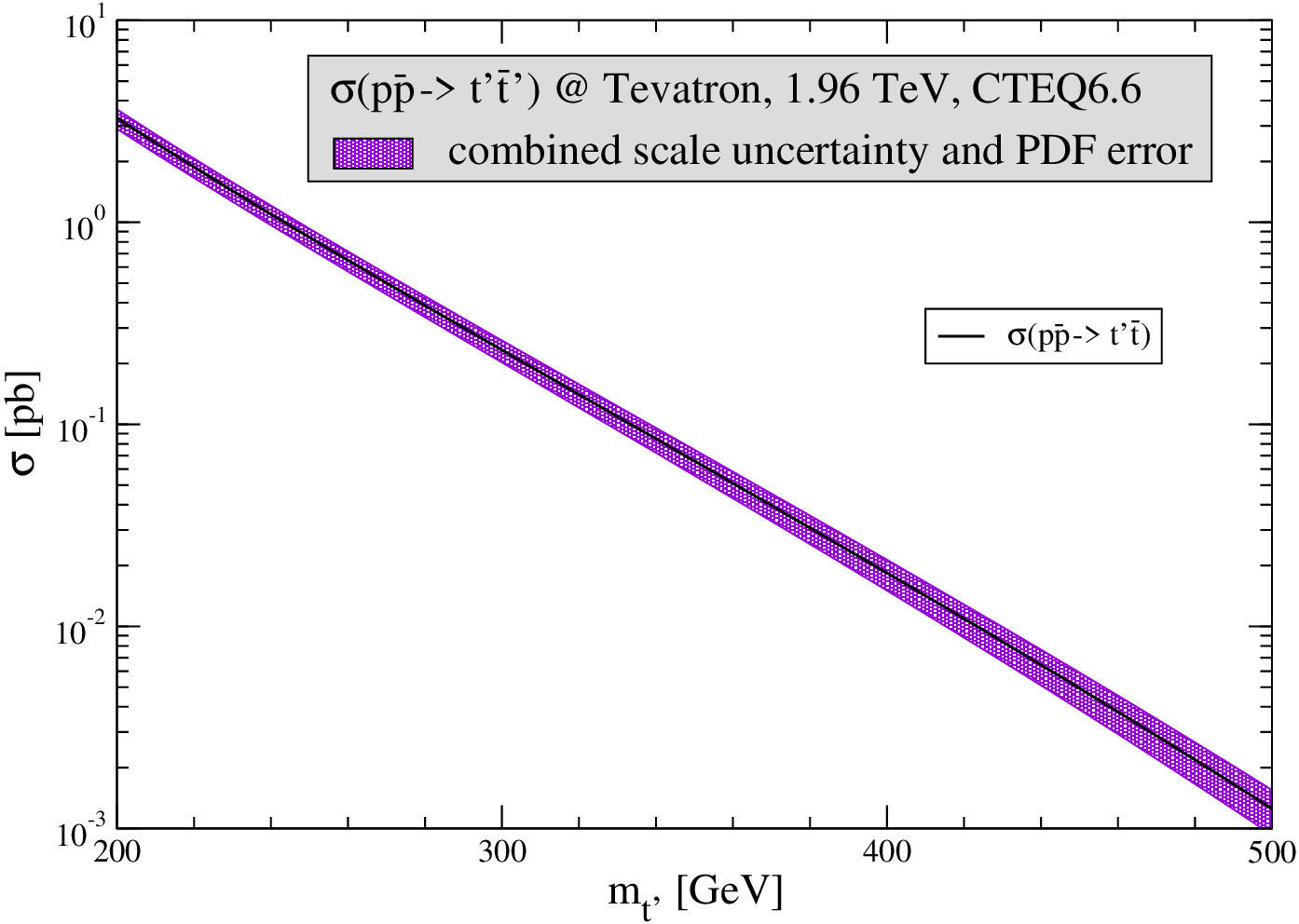}}
\end{minipage}
\begin{minipage}{69mm}
\centering
\hspace*{5mm}
\scalebox{0.28}{\includegraphics{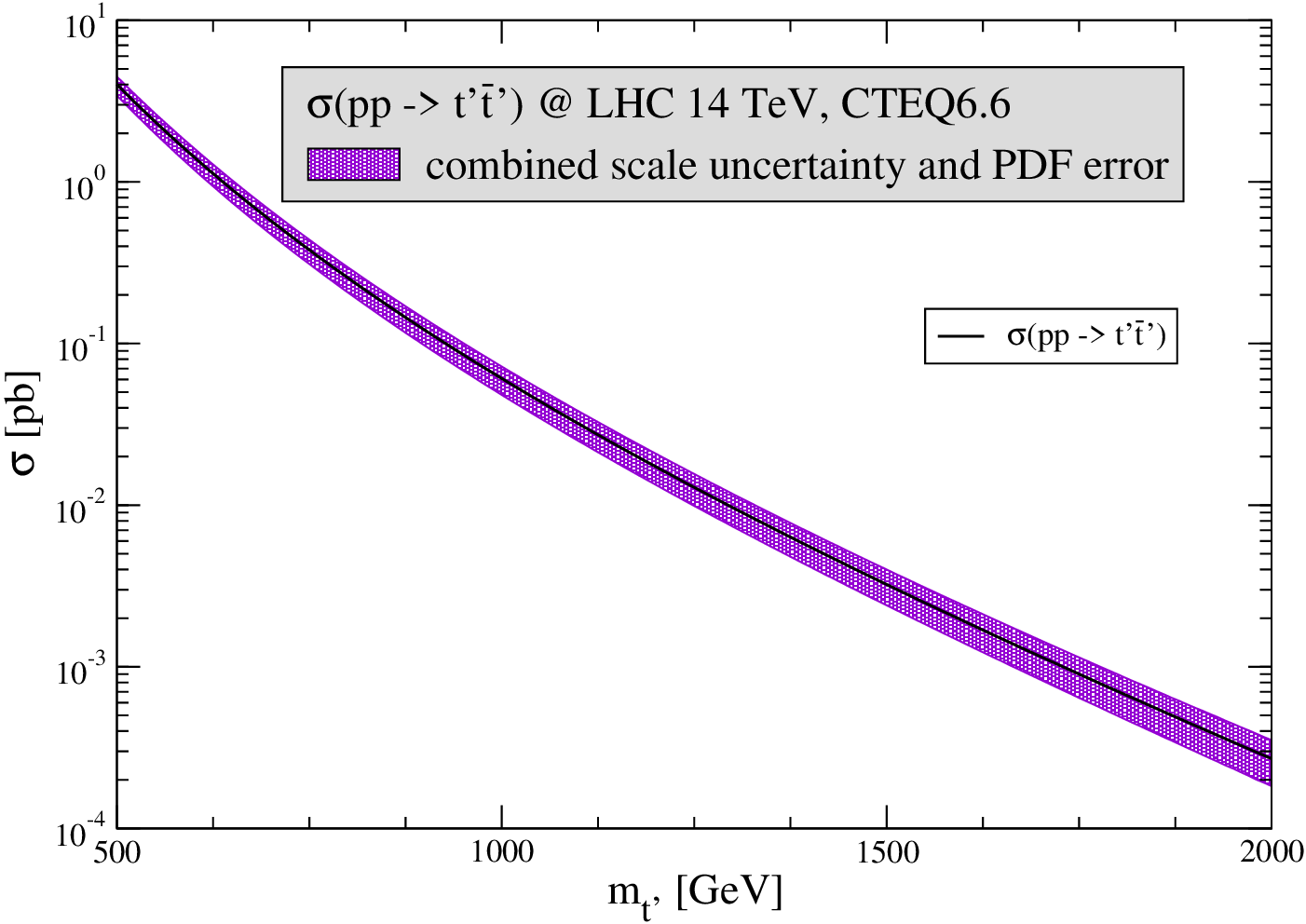}} 
\end{minipage}

\begin{minipage}{65mm}
 \centering
\hspace*{-5mm}
{\caption{
    Pair production cross section for a hypothetical heavy fourth
    generation quark at the Tevatron. The violet band indicates the combined 
    scale uncertainty and PDF error.}
\label{fig:teva:combinederror}}
\end{minipage}
\hspace*{10mm}
\begin{minipage}{65mm}
 \centering
{\caption{Pair production cross section for a hypothetical heavy fourth
generation quark at the LHC. The violet band indicates the combined scale uncertainty 
and PDF error.}
\label{fig:lhc:combinederror}}
\end{minipage}
\end{figure}

We briefly present theoretical predictions for the pair production cross section 
of a hypothetical heavy fourth generation quark $t'$ at the Tevatron and the LHC.
In this calculation we have set the number of light flavours to $n_f = 6$.
As one can see in Figs.~\ref{fig:teva:combinederror} and~\ref{fig:lhc:combinederror}
the cross section decreases very rapidly with increasing $t'$ mass.
At the Tevatron, we predict for a $200\,\GeV$ $t'$ quark a cross section
of $\sigma(p\bar{p}\to t'\bar{t}') = 3.3 \pm 0.3\,\pb$ 
and for $m_{t'} = 500\,\GeV$,
we predict $\sigma(p\bar{p}\to t'\bar{t}') = 1.3^{+0.2}_{-0.4}\,\fb$.
Scale uncertainty and PDF error contribute roughly equal parts to the total error.
At the LHC, we can test higher $m_{t'}$ masses.
We predict for $m_{t'} = 500\,\GeV$ a cross section of 
$\sigma(pp\to t'\bar{t}') = 4.0^{+0.5}_{-0.6}\,\pb$
and for $m_{t'} = 2000\,\GeV$, we have 
$\sigma(pp\to t'\bar{t}') = 0.27^{+0.8}_{-0.9}\,\fb$.
At the LHC, the PDF error is much larger than the scale uncertainty.
Most $t'\bar{t}'$ pairs are produced via the $gg$ channel,
the PDF error of the gluon PDF is large in the relevant kinematic region,
i.e.~at high $x$.
The hadronic cross sections for $t'\bar{t}'$ production including the total error bands
are presented in Fig.~\ref{fig:teva:combinederror} for the Tevatron and in
Fig.~\ref{fig:lhc:combinederror} for the LHC.

\begin{figure}
\begin{minipage}{69mm}
\centering
\hspace*{-5mm}
\scalebox{0.28}{\includegraphics{langenfeld_ulrich_fig3.eps}}
{\parbox{70mm}{\caption{NNLO $t\bar{t}$ production cross section at the Tevatron.
                        The blue band indicates the scale uncertainty.}
\label{fig:teva:massdep}}}
\vspace*{10mm}
\end{minipage}
\begin{minipage}{69mm}
\centering
\hspace*{5mm}
\scalebox{0.28}{\includegraphics{langenfeld_ulrich_fig4.eps}} 
{\parbox{65mm}{\hspace*{10mm}\parbox{64mm}
{\caption{NNLO $t\bar{t}$ production cross section at the LHC. 
          The blue band indicates the scale uncertainty.}
\label{fig:lhc:massdep}}}}
\vspace*{10mm}
\end{minipage}
\begin{minipage}{69mm}
\centering
\hspace*{-5mm}
\scalebox{0.28}{\includegraphics{langenfeld_ulrich_fig5.eps}}
{\vspace*{-2mm}\parbox{68mm}{\caption{Combined scale uncertainty
                       and PDF error for $t\bar{t}$ production
                       at NLO (blue band) and NNLO (red band) 
                       at the Tevatron.}
\label{fig:teva:pdferrormstw}}}
\end{minipage}
\begin{minipage}{69mm}
\centering
\hspace*{5mm}
\scalebox{0.28}{\includegraphics{langenfeld_ulrich_fig6.eps}} 
{\parbox{65mm}{\hspace*{10mm}\parbox{64mm}
{\caption{Combined scale uncertainty and PDF 
                    error for $t\bar{t}$ production  
                    at NLO (blue band) and NNLO (red band) at the LHC.}
\label{fig:lhc:pdferrormstw}}}}
\end{minipage}
\end{figure}

\section*{Acknowledgments}
This work is supported by the Helmholtz Gemeinschaft under contract VH-NG-105 
and by the Deutsche Forschungsgemeinschaft under contract SFB/TR 9.
P.U. acknowledges the support of the Initiative and
Networking Fund of the Helmholtz Gemeinschaft, contract HA-101
("Physics at the Terascale").


\begin{footnotesize}


\end{footnotesize}


\end{document}